\documentstyle[11pt,paspconf,epsf]{article}

\begin{document}

\title{RX J0719.2+6557: A New Eclipsing Polar}

\author{G. Tovmassian (UNAM, Mexico), J. Greiner, F.-J. Zickgraf 
(Garching, Germany), P. Kroll (Sonneberg, Germany), J. Krautter, I.~Thiering 
(Heidelberg, Germany) and A. Serrano (INAOE, Mexico)}

\begin{abstract}
A new magnetic, eclipsing cataclysmic variable is identified as the 
counterpart of the
X-ray source RX J0719.2+6557 detected during the ROSAT all-sky survey. 
The relative phasing of photometric and spectroscopic 
periods indicates a self-eclipsing system.  Doppler
tomography points to the heated surface of the secondary as a strong source 
of emission and of diskless accretion. 
Near-infrared spectroscopy revealed two unusual strong emission features
originating on the heated side of the secondary.
\end{abstract}


Within a program for optical identification of a complete sample of northern 
ROSAT All-Sky Survey sources we identified a new cataclysmic variable.
The X-ray source RX\,J0719.2+6557 
(= 1RXS J071913.4+655734) is detected with 
a total of 66 photons, which corresponds to a vignetting corrected
countrate of 0.16 cts/s. No strong variability in the X-ray intensity is 
seen at this level.

Optical CCD photometry (Johnson B filter) was performed on 
15, 16 and 17 of April 96
at Sonneberg Observatory and 
clearly reveals an eclipsing lightcurve with the 
following orbital elements: 
Min. = HJD 2450189.46326 + 0\hbox{$.\!\!^{\rm d}$}0682297~$\times$~E.
The eclipse FWHM is about 6\,min.
The light curve of the system in between eclipses is rather flat in comparison 
to other self-eclipsing polars (Stockman 1995, Schwope et al. 1995).

Optical spectroscopy was performed at the 2.1\,m telescope of the 
OAN SPM in  April 1996 in two wavelength ranges:
$\lambda\lambda 3600 - 5400 \AA$ with $ 4 \AA$ FWHM re\-so\-lution 
and $\lambda\lambda 6000 - 9000 \AA$ with 
 $ 6 \AA$ resolution. 
The typical Balmer lines accompanied with strong He I and He\,II lines 
in emission strongly suggests a  magnetic nature of this CV. 
The 98.2 min.  period derived from 
the moments of eclipses coincides with the spectroscopic period  within error 
limits.
Based on the large semi-amplitude of the RV curves and their relative 
narrowness for an eclipsing system,  we conclude that the emission lines come 
mostly from the XUV illuminated hemisphere of the secondary star.
Thus, phase 0 (--/+\,crossing of RV curve)  corresponds to the inferior 
conjunction of the secondary. The eclipse occurs at phase 0.4 and is seen also 
in the continuum when the accretion shock disappears partially or entirely  
behind the WD. 
The nature of the X-ray spectrum, the synchronized WD spin and orbital period, 
define RX J0719.2+6557 as a self-eclipsing polar.

\begin{figure}
\vspace{0.0 truecm}
\begin{center}
\setlength{\epsfxsize}{3.7in}\parbox{\epsfxsize}{\epsfbox{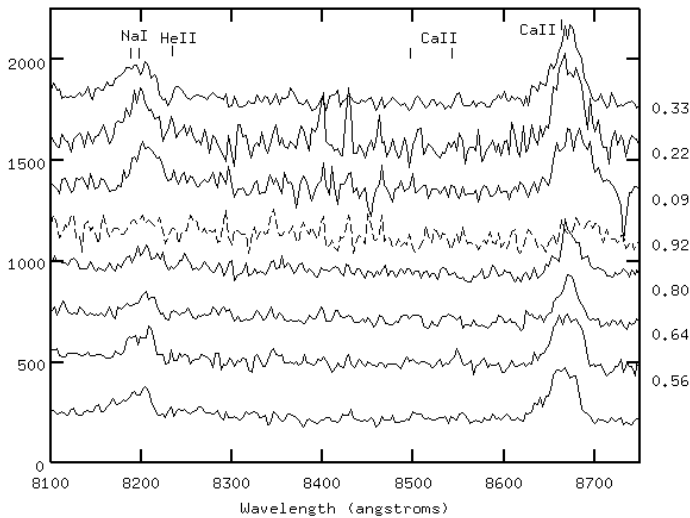}}
\end{center} 
\caption{Spectra of RX\,J0719.2+6557 in the red region. 
The orbital phases are marked on the right side. } \label{fig-1}
\end{figure}

The near-infrared spectra (Figure~\ref {fig-1}) reveal two strong features 
at $\lambda 8200\,\AA$ and $\lambda 8660\,\AA$. These coincide with the 
location of the Na\,I doublet and the component of the Ca\,II triplet. 
These lines are seen as absorption features seated on tremendous broad 
emission.  
Friend et al. (1988) did not find any such emission in a large sample of CVs,
although similar but much fainter features showed up in some accretion disc 
systems after subtraction of the secondary's spectra. These were referred to as 
``disc component" by Friend et al. (1990). In our
 case these lines originate on the leading side of the secondary or on the 
line of sight between the two components, because they both significantly 
decrease during the inferior conjunction of the secondary (phase 0.0). 

The backprojection method (Horne 1991) was applied on the lines of He\,II and 
$H_{\beta}$ with the phase registration adopted above. It is evident that 
the He\,II emission region is compact and concentrated at the expected location 
of the secondary, with a 
possible motion along an horizontal stream. Unlike He\,II, $H_{\beta}$ shows 
more diffuse distribution around secondary and the inner $L_1$ point. 
There is also some evidence of a vertical stream in the light of $H_{\beta}$. 
Such feature was observed in a similar system (Schwope et al. 1995), but its 
nature is not  clear. Although we have fewer spectra in the red, it is 
obvious that the emission features in this region are  strongly concentrated 
on the leading side of the secondary or along the vertical stream observed 
in $H_{\beta}$.

\end{document}